# Yanson point-contact spectroscopy of Weyl semimetal WTe$_2$.

Yu. G. Naidyuk[1], D. L. Bashlakov[1], O.E. Kvitnitskaya[1], S. Aswartham[2], I. V. Morozov[2,3], I. O. Chernyavskii[2,3], G. Shipunov[2], G. Fuchs[4], S.-L. Drechsler[5], R. Hühne[4], K. Nielsch[4], B. Büchner[2], D. V. Efremov[5]

[1]*B. Verkin Institute for Low Temperature Physics and Engineering, NAS of Ukraine, 61103 Kharkiv, Ukraine*

[2]*Institute for Solid State Research, IFW Dresden, D-01171 Dresden, Germany*

[3]*Lomonosov Moscow State University, Moscow, 119991, Russian Federation*

[4]*Institute for Metallic Materials, IFW Dresden, D-01171 Dresden, Germany*

[5]*Institute for Theoretical Solid State Physics, IFW Dresden, D-01171 Dresden, Germany*

We carried out point contact (PC) investigation of WTe$_2$ single crystals. We measured Yanson PC spectra ($d^2V/dI^2$) of the electron-phonon interaction (EPI) in WTe$_2$. The PC spectra demonstrate a main phonon peak around 8 meV and a shallow second maximum near 16 meV. Their position is in line with the calculation of the EPI spectra of WTe$_2$ in the literature, albeit phonons with higher energy are not resolved in our PC spectra. An additional contribution to the spectra is present above the phonon energy, what may be connected with the peculiar electronic band structure and need to be clarified. We detected tiny superconducting features in $d^2V/dI^2$ close to zero bias, which broadens by increasing temperature and blurs above 6K. Thus, (surface) superconductivity may exist in WTe$_2$ with a topologically nontrivial state. We found a broad maximum in $dV/dI$ at large voltages (>200 mV) indicating change of conductivity from metallic to semiconducting type. The latter might be induced by the high current density (~$10^8$ A/cm$^2$) in the PC and/or local heating, thus enabling the manipulation of the quantum electronic states at the interface in the PC core.

## Introduction

The rich family of transition metal dichalcogenides (TMDs) exhibits many interesting physical properties such as the metal/insulator transition, semiconducting and semimetallic transport properties, magnetism, charge-density wave state and superconductivity. Most of these properties steam from the strongly two-dimensional character associated with the layered crystal structure by all of them. The effect of the two-dimensional character of these compounds on the physical properties was already noticed by Wilson & Yoffe in their comprehensive review in 1969 [Wilson]. In this review they listed more than 60 compounds with their physical properties and pointed out that at least two third are layered compounds. TMDs crystallize in different structures denoted as 2H-, $T_d$-, 1T-, and 1T`-type lattices. The 2H- and 1T-type compounds are primarily semiconducting, whereas the $T_d$- and 1T`- type compounds are predominantly semimetallic.

Recently, enormous interest in layered TMDs arose after discovery of huge non-saturating magnetoresistance in $WTe_2$ [Ali] and especially after prediction in Ref. [Sol] that some TMDs, e.g. $WTe_2$ and $MoTe_2$, may demonstrate novel topological quantum properties connected with formation of low-energy electronic excitations, the so-called Weyl fermions with linear dispersion along all the three momentum directions. Specifically, Weyl semimetals obey peculiar band structure both in the bulk and on the surface, what is responsible for their expected breathtaking properties, with promising prospects for low dissipation quantum electronics and spintronics [Yan].

In addition to the nontrivial normal state, $WTe_2$ and $MoTe_2$ display transition to the superconducting (SC) state under pressure. The critical temperature $T_c$ in $WTe_2$ increases strongly under pressure reaching a maximum of around 7 K in the range of 13–17 GPa [Pan, Kang]. Bulk superconductivity and topological edge states at the surface cause genuine interest in these materials from the point of view of studying exotic topological superconductivity.

In our recent paper [Naid18], we have investigated $MoTe_2$ by point-contact (PC) spectroscopy. We have discovered a tremendous enhancement of the SC critical temperature up to 5K in $MoTe_2$ at the surface seen by the surface sensitive PC measurements. The origin of the surface superconductivity was assigned to the topological surface states, since $MoTe_2$ is the Weyl semimetal of type II. Here, we report of a similar study of the sister compound $WTe_2$ in the normal and SC state. Applying of Yanson PC spectroscopy in the normal state [Naid_book], we obtained the electron-phonon interaction (EPI) in $WTe_2$, which is key element in the development of the microscopic theory of the superconductivity in this compound. Additionally, we found shallow SC features in PC characteristics, which did not allow obtaining detailed information about the SC state, viz. the SC gap, similar to that $\alpha^2F(\omega)$ found in $MoTe_2$ [Naid18]. Thus, the results as to EPI and SC state in $WTe_2$ turned up quite different from that obtained for $MoTe_2$. Furthermore, we have observed the transition from metallic to semiconducting type of conductivity in PC at a high current density.

## Experimental details and results

Bulk single crystals of $WTe_2$ were grown with Te flux. To avoid contamination, the mixing and weighting were carried out in an Ar-filled glove box. Amounts of 0.5 g of W powder and 10g Te were mixed and placed in an evacuated quartz ampule. The ampule was placed in a box furnace and slowly heated to 1000 °C and cooled down slowly to 800 °C followed by a hot centrifuge to remove the excess Te-flux. Single crystals were grown having a needle-like shape with a layered morphology. The as-grown crystals were characterized by SEM in EDX mode for compositional analysis and by x-ray diffraction for structural analysis

PCs were prepared by touching of a thin Ag wire to a cleaved at room temperature flat surface of a $WTe_2$ flake or contacting the edge of a plate-like sample by this wire. Additionally, we prepared several so-called "soft" PCs by dripping of a small drop of silver paint onto the cleaved $WTe_2$ surface. The latter type of PCs can prevent any pressure effect which is possible at the former "mechanical or "hard" PC preparation. Furthermore, the "soft" PCs demonstrate better stability. Thus, we conducted resistive measurements of heterocontacts between a normal metal (Ag or silver paint) and the WTe$_2$.

We measured current-voltage characteristics $I(V)$ of PCs and their first and second derivatives. The first derivative or differential resistance $dV/dI(V)\equiv R(V)$ and the second derivative $d^2V/dI^2(V)$ or Yanson PC spectrum were recorded by sweeping the dc current $I$ on which a small ac current $i$ was superimposed using a standard lock-in technique. The measurements were performed mainly in the temperature range of liquid helium and in magnetic field.

We have measured in total 20 PCs with resolved EPI spectral features. Among them were 18 WTe$_2$–Ag heterocontacts made by touching of a thin Ag wire to the edge of a WTe$_2$ ribbon (flake) or to its flat surface. Also, a number of "soft" PCs were prepared by putting a small drop of silver paint on the flat surface or edge of a WTe$_2$ ribbon. Among the latter, two PCs showed spectral EPI maxima.

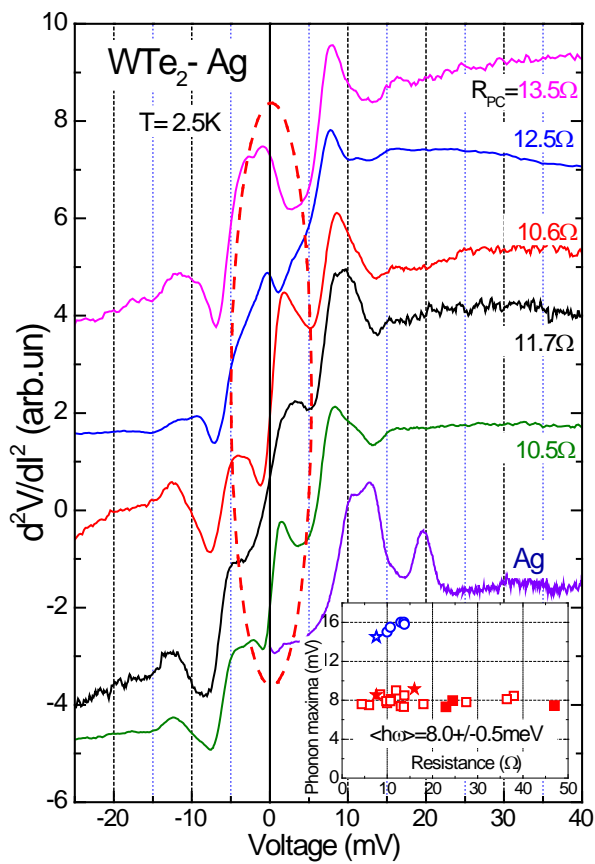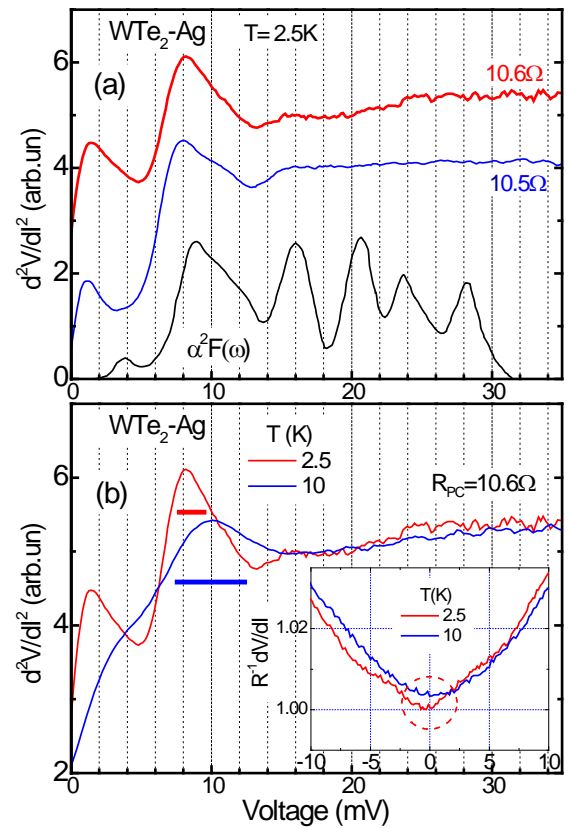

Fig. 1. Series of PC $d^2V/dI^2$ spectra of several WTe$_2$–Ag PCs with a normal state resistance shown by each curve. Bottom (blue) curve is the PC $d^2V/dI^2$ spectrum of clean Ag [Naid82]. Zero bias features inside the dashed oval are of non-phononic nature (see text). The inset shows the position of the main phonon peaks in $d^2V/dI^2$ for all measured PCs: open symbols – PCs made by touching of Ag wire to the edge of WTe$_2$ ribbon, solid symbols - PCs made by touching of Ag wire to the flat surface of WTe$_2$ ribbon, stars – 2 "soft" PCs made with silver paint.

Fig. 2. a) Comparison of two $d^2V/dI^2$ spectra from Fig.1 with calculated EPI function $\alpha^2F(\omega)$ of 1T`-WTe$_2$ at 10 GPa [Lu]. b) Smearing of $d^2V/dI^2$ spectrum with increasing temperature up to 10K. Horizontal bars below the main maximum show the resolution for each spectrum, which is temperature dependent. The inset shows the differential resistance for the curves from the main panel. A shallow SC minimum (indicated by dashed circle) develops at low temperature of 2K.

Fig.1 shows measured $d^2V/dI^2$ curves with spectral features (maxima) for several WTe$_2$–Ag PCs. According to the Yanson PC spectroscopy [Naid_book], $d^2V/dI^2$ of ballistic (or diffusive) PCs reflects the processes of conduction electron interaction with phonons or other quasiparticles, which reduce the conductivity of the PCs. Here, we see a main maximum at around 8mV and more shallow second one near 16 mV, which we associate with the EPI. The inset in Fig.1 shows the position of these maxima in $d^2V/dI^2$ for all measured PCs. Most of the heterocontacts were prepared by touching of an Ag wire to the edge of a WTe$_2$ ribbon (flake), that is presumably along the ab-direction. A number of PCs were prepared by touching of an Ag wire to the WTe$_2$ plane surface, that is, presumably along the c-direction. However, no principle difference in the main peak position (Fig.1, inset) and their relative intensity is seen. The position of the main maximum averaged for 20 PCs spectra is 8.0+/-0.5 mV.

For the three bottom curves in Fig.1, a sharp peak close to zero bias is detected. This peak is suppressed by increasing the temperature (see Fig. 2b) or by a magnetic field. Thus, it can be assigned to residual superconductivity in the PC. The bottom $d^2V/dI^2$ curve in Fig.1 is the PC EPI spectrum of Ag [Naid82], with the main maxima located between 10.5 mV and 13 mV related to transverse double maxima peaks of acoustic phonons and at 20 mV for longitudinal acoustic phonons. As can be seen, the main phonon modes of the Ag tip (which is used as counterelectrode) do not give a discernible contribution to the $d^2V/dI^2$ spectra of the WTe$_2$–Ag PC. Therefore, all features in the spectra are caused by the WTe$_2$ phonons.

In Fig. 2a, two $d^2V/dI^2$ PC spectra from Fig.1 are compared with the calculated function of EPI in WTe$_2$ at a pressure of 10 GPa[1] [Lu]. The comparison of the measured and calculated spectra confirms that $d^2V/dI^2$ reflects the EPI function in WTe$_2$, albeit the phonons with energies above 16 meV are not resolved in our PC spectra. Note, that three bottom curves in Fig. 1 display the SC maximum close to zero bias, which corresponds to a zero-bias minimum in $dV/dI$ (see Fig.2b, inset). In contrast, the two upper curves in Fig. 1 show a so-called negative anomaly (minimum) corresponding to a zero-bias maximum in $dV/dI$. Such anomaly is usually due to the scattering on magnetic impurities (viz. Kondo effect) [Naid82mn] or on two-level excitations caused by local disorder [Keijser]. Thus, the latter spectra demonstrate the absence of superconductivity in the PC core. Furthermore, these spectra display the main phonon peak at 7.5 mV, while the bottom three spectra with their close to zero-bias SC maximum show this peak at around 8 mV or a bit higher.

We succeed to measure $d^2V/dI^2$ PC spectra with EPI structure also for two "soft" PCs. In general, they show similar phonon maxima at low energies (see Fig.3.). Besides, the PC spectra intensity above the phonon structure falls down to a minimum at 50-60 mV with a subsequent background increase (see bottom inset in Fig.3a and main panel of Fig.3b). "Soft" PCs are quite stable, that allowed us to measure spectra at larger voltages and at higher temperatures. $d^2V/dI^2$ PC spectra at two temperatures are shown in the insets of Fig.3, while the corresponding first derivatives $dV/dI$ are shown in Fig.4 for several temperatures. The remarkable feature in $dV/dI$ is an apparent maximum above 200 mV, which shifts towards the lower voltages and smears with increasing temperature.

## Discussion

We have measured EPI spectra of WTe$_2$ using Yanson PC spectroscopy. The comparison with the calculated EPI spectrum for this compound (Fig.2a) gives evidence that the measured spectra reflect the EPI function in WTe$_2$, albeit phonons with energy higher than 16 meV are not resolved in our PC spectra. The latter may be due to enhancement of the inelastic scattering with increasing of excess energy of electrons with the bias rise and transition to the nonspectral (thermal) regime [Verkin, Naid_book] (see more details below). It's interesting at this point that the Debye temperature in WTe$_2$ is estimated to be about 100 K (~10 meV) and the Fermi energy is within 17−32 meV [Rana]. The low Debye energy may indicate that indeed the high energy phonons give only a small contribution to the thermodynamics and likely to the electronic transport. On the other hand, a tiny Fermi energy, which is comparable to the Debye energy, can restrict scattering on the phonons with higher energy. Here, it should be noted, that the calculated spectrum in Fig.2a represents the *thermodynamic* EPI function, while the PC spectrum itself is some kind of transport EPI function [Naid_book], which underlines large angle electron scattering leading to different electron-phonon coupling constants.

---

[1] According to Ref. [Kang], the superconductivity begins to appear in WTe$_2$ at this pressure. At the same time, according to Ref. [Pan], T$_c$ reaches already several Kelvin at 10 GPa. That is, the superconductivity in WTe$_2$ is sample dependent and external impact sensitive.

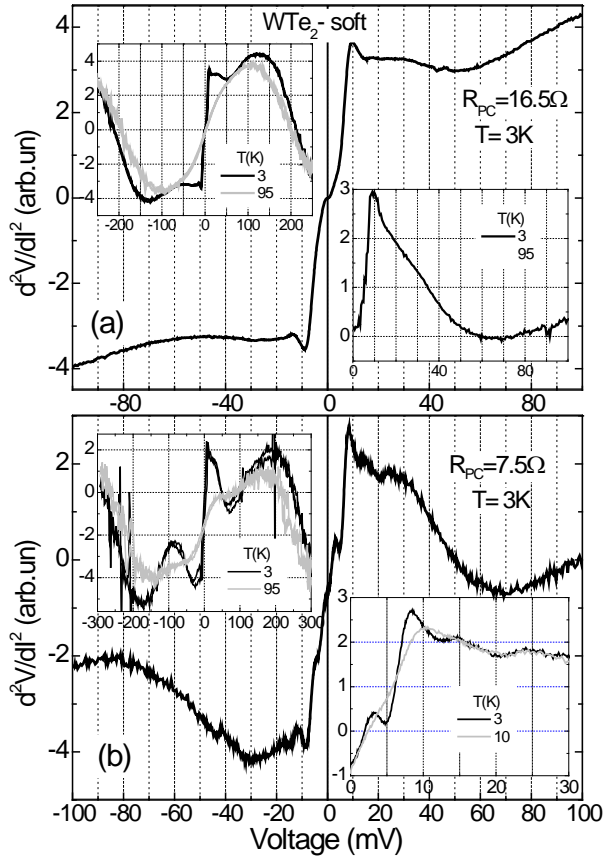
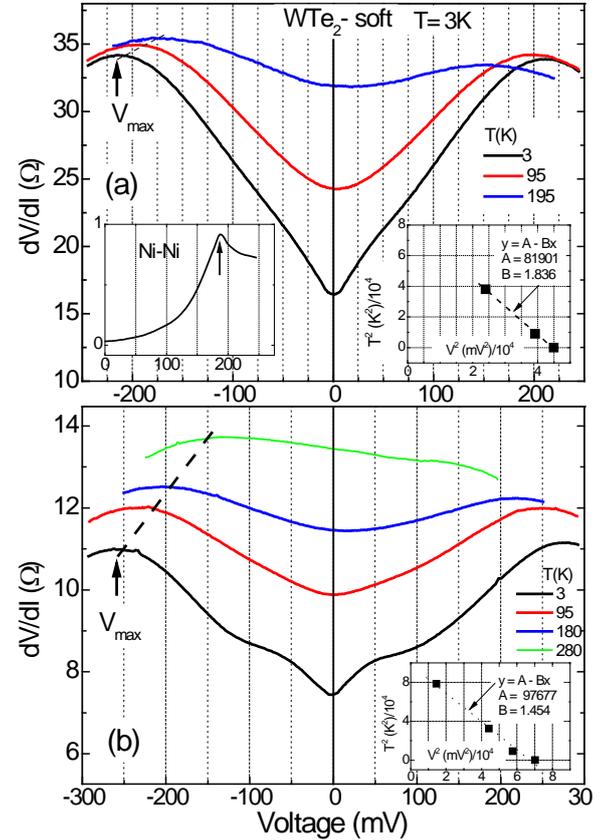

Fig. 3. $d^2V/dI^2$ spectra of two soft WTe$_2$ PCs with a normal state resistance $R_{PC}$=16.5Ω and 7.5Ω. Upper insets in each panel show $d^2V/dI^2$ spectra at T=3K and 95K for a larger bias. Bottom inset in panel (a) shows the difference between $d^2V/dI^2$ at 3K and 95K from the upper inset. Bottom inset in panel (b) shows $d^2V/dI^2$ measured at 3K and 10K in the range of phonon energies.

Fig. 4. $dV/dI$ curves for PCs from Fig. 3 at larger bias and for different temperatures. Left inset in panel (a) shows $dV/dI$ for nickel PC [Verkin] with maximum around 200 mV. Right insets show the dependence of the maxima position $V^2_{max}$ averaged for two polarities versus $T^2$.

A number of PCs demonstrate spectra with tiny SC features like a maximum around 1-2mV, which smears out with the temperature rise above 6K. These spectra have the main phonon peak position about 0.5-1 mV higher than that for spectra without signature of superconductivity. This shift of the phonon peak to a higher energy or the hardening of the phonon modes is commonly connected with a pressure effect. Thus, we assume that the SC features in the PC spectra of WTe$_2$ can appear due to the local pressure caused by the Ag counter-electrode at the PC formation. This would also explain why the SC structure in MoTe$_2$ PCs is much more pronounced than in WTe$_2$ PCs taking into account the much stronger effect of pressure on $T_c$ in MoTe$_2$[2]. At the same time, we observed the similar feature for "soft" PC (see Fig. 3b and bottom inset), where a pressure effect is unlikely. This might point to surface superconductivity. Here, we should also mention that $T_c$ of WTe$_2$ reaches up to 1.6K in epitaxial thin film with thickness of 10 nm (7 unit cells) [Asaba]. It looks like that the superconductivity in WTe$_2$ is possible stimulate not only by the pressure effect. According to Ref. [Lu], the superconductivity in WTe$_2$ under pressure emerges from the monoclinic 1T` phase rather than from the orthorhombic T$_d$ phase under ambient conditions. However, Asaba *et. al.* [Asaba] concluded that their SC epitaxial thin films have still orthorhombic T$_d$ structure.

Let us compare the results for WTe$_2$ with our previous PC measurements in the sister compound MoTe$_2$ [Naid18]. Contrary to the present measurements, the SC features in *dV/dI* of MoTe$_2$–Ag PCs were much more pronounced and they even showed Andreev-like structure that enabled us to obtain the SC gap value and its temperature and magnetic field dependence. On the other hand, we measured more detailed EPI spectra in the case of WTe$_2$ with good reproducibility, while EPI spectra in MoTe$_2$ were rare and showed only a broad maximum between 15-20 mV [Naid18]. At the same time, the calculated phonon density of states in MoTe$_2$ has a rich structure [Tse] and, on the whole, has some similarity to that of WTe$_2$ [Lu] below 20 meV. These issues require further studies.

No other features similar to quasiparticle excitations are seen in the measured spectra, what is in line with the conclusion from Ref. [Lu] that the calculated EPI under pressure predicts a similar $T_c$ as the reported value, implying that WTe$_2$ might belong to the class of conventional BCS superconductors. However, the behavior of $d^2V/dI^2$ spectra of "soft" PCs above the phonon maxima is peculiar and probably of non-phonon origin if the rare case of multiphonon processes due to strong anharmonicity is excluded. As seen in the spectrum shown in Fig.3b, a broad falling background is seen up to about 60 mV. A similar behavior shows also the spectrum in Fig.3a, especially after subtracting from the spectrum the data measured at 95K[3] (see upper inset in Fig. 3a). The reason for this unusual challenging behavior is not yet clear and will be addressed elsewhere. Probably, it is due to the small Fermi energy and the specific band structure of WTe$_2$ and/or related to specific excitations related to its topological surface state.

The use of the "soft" PCs allows us to explore the spectra at large voltages and at high temperatures. This way, the upturn of *dV/dI* from the metallic to the semiconducting behaviour is found, which produces a broad maximum above 200 mV (Fig.4). The latter is shifted towards the lower voltages and smears with increasing temperature disappearing close to the room temperature. A maximum in *dV/dI* at biases well above the Debye energy was first observed in PCs with ferromagnetic metals [Verkin, Naid_book]. It was attributed to the realization of the thermal regime, when the temperature in the PC increases with the bias and reaches the Curie temperature. A kink in resistivity occurs at the phase transition from the ferromagnetic to the paramagnetic state at this temperature, which produces a maximum in *dV/dI*. According to the theory of the thermal regime [Verkin, Naid_book], the temperature $T_{PC}$ in PC increases with the voltage as:

$$T_{PC}^2 = T_{bath}^2 + V^2/4L_0, \quad (1)$$

where $T_{bath}$ denotes the temperature of the bath and $L_0 = 0.0245$ mV$^2$/K is the Lorenz number. If we fix $T_{PC}$ to the characteristic temperature at which some peculiarity in the resistivity occurs, then the voltage at which this peculiarity appears in *dV/dI* will shift with the $T_{bath}$ increase according to Eq. (1) that is "linearly" in the coordinates $V^2$ versus $T_{bath}^2$. Such a behavior is shown in the right insets in Fig. 4. From the linear fit of $V^2$ versus $T_{bath}^2$, the Lorenz number and the characteristic temperature can be estimated. These values are in the range of 5-7$L_0$[4] and 280-320K for the PCs from Fig.4. The increased value of the Lorenz number is in line with the thermal conductivity measurements of WTe$_2$ [Mleczko], where the phonon parts dominate while the electronic contribution is only within 10-30%. Our estimated characteristic temperature of about 300 K is not far from the position of the broad resistivity maximum around 400K in WTe$_2$ [Kaba, Jana]. The latter is connected, according to Ref. [Jana], with the thermal excitation of carriers across the indirect narrow band gap near the Γ-point. Thus, the behavior of *dV/dI* at high voltages can be explained by the realization of the thermal regime due to shortening of the inelastic mean free path of energized electrons with the bias raised.

---

[2] According to [Pan], the superconductivity sharply appears in WTe$_2$ at a pressure of 2.5 GPa, while $T_c$ in MoTe$_2$ reaches already 6K at this pressure [Qi].

[3] In general, the difference between the EPI spectra measured at different temperatures comes from the resolution. That is, a sharp spectral maximum will broaden by temperature into bell-shaped line with the width 5.44$k_B$T [Naid_book], which is 1.4meV at 3K and 45 meV at 95K. Therefore, all phonon maxima are smeared out in the spectra at 95K.

The temperature dependence of the PC resistance at zero bias can be used for a more reliable determination of the PC size and other parameters (see Supplement). The size (diameter) of the PCs from Figs. 3 or 4 is estimated as 125 nm and 400 nm for the PCs from panel (a) and (b), respectively. As a result, the current density reaches a value of about $10^8$ A/cm$^2$ at the position of the maximum in *dV/dI*. So, WTe$_2$ PCs can carry currents with a very high density, which is even higher than $5 \cdot 10^7$ A/cm$^2$ reported for ultrathin (3−20 layers thick) WTe$_2$ films [Mlleczko].

## Conclusion

Yanson PC spectra demonstrate the EPI structure of WTe$_2$ with the main phonon peak around 8 meV and a shallow second maximum near 16 meV. Another type of quasiparticle interaction is not seen in the spectra, what indicates that EPI is the main glue for the Cooper pairing in this compound. At the same time, an additional contribution to the Yanson spectra is present above the phonon energy, what may be connected with the peculiar electronic band structure, specific excitations for the topological surface system realized in WTe$_2$. Each of these unusual challenging options needs to be clarified in future experimental and theoretical studies.

We have detected tiny SC features in the *d²V/dI²* spectra like a peak close to zero bias, which broadens by increasing the temperature and blurs above 6K. The corresponding SC zero-bias minimum in *dV/dI* was very shallow that is well below 1% of the PC resistance, contrary to the much more pronounced SC structure in *dV/dI* measured for the sister MoTe$_2$ compound in our previous investigations. The origin of the SC feature may be both the pressure effect for "hard" PCs and surface superconductivity of possibly different topological nature or symmetry.

We have shown that WTe$_2$ PCs with submicron size are able to withstand ultrahigh current density of about $10^8$ A/cm$^2$ and can change their conductivity from metallic to semiconducting type under high bias or high current density. This is another advantage of nanosized WTe$_2$, since this material is considered to be perspective in wide type of applications such as of low-dissipation electronics, spintronics, optoelectronics, thermodynamics, catalysis, etc [Tian].

Thus, the further study of transport and SC phenomena of WTe$_2$ and related TMDs in restricted geometry like PC is of great interest.

## Acknowledgments

We thank Dr. M. Geier for discussions of topological aspects and K. Nenkov for the technical assistance. This work was financially supported by the Volkswagen Foundation in the frame of Trilateral Initiative. Yu.G.N., D.L.B. and O.E.K. are grateful for support by the National Academy of Sciences of the Ukraine under project Ф4-19 and would like to thank the IFW Dresden for hospitality. I.V.M, S.A., B.B. and D.V.E thank DFG and RSF for financial support in the frame of the joint DFG-RSF project "Weyl and Dirac semimetals and beyond - prediction, synthesis and characterization of new semimetals". S.A., G.S and D.V.E. acknowledge the IFW "Excellence program".

---

[4] We calculated Lorenz number for MoTe$_2$, using data from [Sakai, Supplement] for resistivity and thermal conductivity. It turns out that/$L/L_0$ varies between 4-8 with temperature, which leads to an average of about 7.

**Supplement:**

Determination of the PC parameters.

We use equation for the PC resistance for contacts between two (1 and 2) metals by modifying the Wexler formula [Naid_book] for the case of a heterocontact:

$$R_{het}=8(\rho l_1+\rho l_2)/3\pi d^2 + (\rho_1+\rho_2)/2d \qquad (1)$$

Here, $d$ is the PC diameter, $\rho$ is the resistivity, $l$ is the mean free path of electrons. The product $\rho l$ is constant for a given material (see Eq.(7) below). The first term in (1) is the so-called Sharvin resistance $R_S$ and the second term is the so-called Maxwell resistance $R_M$. Only $R_M$ is temperature dependent because of $\rho = \rho(T)$. By differentiating Eq.(1) over the temperature, we can write down a formula to determine the PC diameter, similar to [Akim]:

$$d=(\partial\rho_1/\partial T +\partial\rho_2/\partial T)/2(\partial R_{het}/\partial T) \qquad (2)$$

This way to determine the PC diameter gives its more precise value, because $\rho_1$ and $\rho_2$ in a PC core can be enhanced compared to their values in the bulk due to imperfect surface and disturbed material in the PC. Usually, a simple clean metal is used as a counter electrode, which resistivity is much smaller than that of the investigated (complex) compound. Therefore, formulas (1) and (2) can be reduced to:

$$R_{het}=8(\rho l_1+\rho l_2)/3\pi d^2 + \rho_1/2d \qquad (3)$$

and

$$d=(\partial\rho_1/\partial T)/2(\partial R_{het}/\partial T) \quad (4)$$

Multiplying numerator and denominator in the second term in Eq.(3) by $l_1$, we receive a formula to determine $l_1$ in a PC:

$$l_1= \rho l_1/2d[R_{het} - 8(\rho l_1+\rho l_2)/3\pi d^2] \quad (5)$$

or, if we consider $\rho l_1 >> \rho l_2$, Eq.(5) transforms to:

$$l_1 \approx \rho l_1/2d(R_{het} - 8\rho l_1/3\pi d^2) \quad (6)$$

The PC resistance was measured for two PCs at several temperatures as shown in Fig. 4. We calculated the averaged $\partial R_{het}/\partial T$ between 3 and 200K for these two PCs with the values 0.08 Ω/K and 0.025 Ω/K, respectively. Then, we used the averaged $\partial\rho_1/\partial T \approx 2\cdot 10^{-6}$ Ω·cm/K in the corresponding temperature range (between 3 and 200K) from the literature [Ali, Pan]. As a result, the calculated diameter is turned out to be about 125 nm for the PC from Fig.4a and 400 nm for the PC from Fig.4b.

To get the $l_1$ value from Eq.(6), we need data for $\rho l_1$. The latter can be estimated from the Drude (free electron) model:

$$\rho l = p_F/ne^2, \quad (7)$$

where $p_F$ is the Fermi momentum, n and e are the electron density and its charge, respectively. Using the relation $p_F = (3\pi^2 \hbar^3 n)^{1/3}$, Eq.(7) transforms to÷ $\rho l \approx 1.28 \cdot 10^4 \cdot n^{-2/3}$ [Ω·cm²], where n is in [cm$^{-3}$].

The electron density in $WTe_2$ varies between $10^{19}$ and $10^{20}$cm$^{-3}$ according to the literature [see, e.g., Li, Kang, Lv, Rana]. We used for the calculation of $\rho l$ value $1.8\cdot 10^{19}$cm$^{-3}$ [Joseph] extracted from Shubnikov-de-Haas oscillations on similar $WTe_2$ samples [Joseph]. Calculation for several other electron density values is given in the attached Table.

Finally, we obtain from Eq.(6) as mean free path of electrons 116 nm and 36 nm for the corresponding PCs, which leads to resistivities in both PCs of $160\cdot 10^{-6}$ Ω·cm and $521\cdot 10^{-6}$ Ω·cm. Furthermore, the relation between the Maxwell and the Sharvin resistance $R_M/R_S$ is 0.63 and 6.6 for these PCs. Thus, we got strongly enhanced resistivities for both PCs with $l \sim d$ for the PC with the smaller resistivity and the diffusive regime $l << d$ for the second one.

It is interesting also, that the current density in these PC reaches a high magnitude. Using the estimated diameter values, we get a current density of about $4\cdot 10^6$ A/cm² and $8\cdot 10^5$ A/cm² for these PCs at 10 mV (e.g., at the position of the first phonon peak), which even increases twentyfold above 200 mV where a maximum in $dV/dI$ develops. This high current density is in accord with the value of $5\cdot 10^7$ A/cm² reported for ultrathin (3−20 layers thick) $WTe_2$ films [Mlleczko].

Of course, our calculations are approximations. E.g., we postulate that the materials in the PC occupy equal parts, including a coefficient 1/2 in the Maxwell part of resistance and similar for the Sharvin resistance, while we must use $\rho l_{(1,2)}$ and $\rho_{(1,2)}$ with a proper (but unknown) weighting factor $w$ ($w_1+w_2=1$), instead of $w_1=w_2=1/2$. Also, as mentioned, the carrier density, which is used to calculate $\rho l$, varies greatly in the literature. Anyway, we believe that the calculated values are more or less reasonable.

Parameters for two PCs calculated for several values of the electron density **n**

| PC from Fig.4a with R=16.5Ω and d=125·10$^{-7}$ cm | | | | | |
|---|---|---|---|---|---|
| **n** ·10$^{19}$ cm$^{-3}$ | $\rho l$ · 10$^{-9}$ Ω·cm² | $\rho$ ·10$^{-6}$ Ω·cm | $l$ · 10$^{-7}$ cm | $l/d$ | $R_M/R_S$ |
| 1 | 2.75 | 38.4 | 714 | 5.71 | 0.1 |
| 3.3 | 1.24 | 243 | 51 | 0.41 | 1.43 |
| 10 | 0.59 | 331 | 18 | 0.14 | 4.06 |
| PC from Fig.4b with R=7.5Ω and d=400·10$^{-7}$ cm | | | | | |
| 1 | 2.75 | 483 | 57 | 0.143 | 4 |
| 3.3 | 1.24 | 547 | 23 | 0.058 | 10 |
| 10 | 0.59 | 575 | 10.3 | 0.026 | 22 |